\documentclass[
    ,final            
  ]
  {aipproc}

\layoutstyle{6x9}
\def\roughly#1{\raise.3ex\hbox{$#1$\kern-.75em\lower1ex\hbox{$\sim$}}}

\begin{document}

\title{High-energy black hole production}


\author{Steven B. Giddings\footnote{Email: giddings@physics.ucsb.edu.}}{
  address={Department of Physics, University of California, Santa Barbara, CA 93106}
}

\begin{abstract}
Black hole production in high-energy collisions is briefly surveyed.  Included is a summary of recent developments and open problems relevant to collider (LHC) production, as well as of some theoretical issues pointing towards fundamental principles of quantum gravity.
 \end{abstract}

\maketitle


\section{Introduction}

Black hole production may be the most spectacular physics at future colliders, perhaps even beginning with the 2008 startup of the LHC.  And even if it is not phenomenologically accessible in the near future, it raises some very profound theoretical issues. This talk will present a summary and update on some of the physics of black hole creation at high energies.\footnote{Due to space limitations only some representative references can be given.  Other reviews include \cite{Kantir,Webber,KBS}.}

The basic idea is that once we reach collision energies exceeding the Planck mass, which we denote $M_D$, collisions of particles can form black holes.  The Planck mass is traditionally expected to be of order $10^{19}$ GeV, but in TeV-scale gravity scenarios the Planck mass could be lowered to the accessible energy $\sim 1$ TeV.  Possible realizations of TeV-scale gravity include large extra dimensions\cite{ADD,AADD} and large warping\cite{RSone,GKP,DoKa}.  If the $D$-dimensional metric is of the general warped form,
\begin{equation}
ds^2 = e^{2A(y)} dx_4^2 + g_{mn}(y)dy^m dy^n\ ,
\end{equation}
where $y^m$ are coordinates for the extra dimensions and $e^{2A}$ is the warp factor, 
the basic relation between the four-dimensional Planck scale and the $D$-dimensional Planck scale is
\begin{equation}
M_4^2 = M_D^{D-2} \int {d^{D-4} y\over (2\pi)^{D-4}} \sqrt{g_{D-4}} e^{2A}\ .
\end{equation}  
If the integral on the RHS of the equation, called the ``warped volume," is large, this can yield the measured four-dimensional gravitational constant with an $M_D$ significantly below $10^{19} GeV$.  In the large extra dimensions case, this is readily understood as due to  dilution of the field lines in the extra dimensions.  One must supplement this picture with a brane-world scenario in order that the ordinary gauge forces behave four-dimensionally.  Many such scenarios are being investigated, in string theory and elsewhere.  

Once it became clear that the true Planck scale could be as low as a TeV, it was obvious that black hole production could take place at this scale.  Early discussions of this in large extra dimension scenarios and warped scenarios are \cite{BaFi,GiKa} respectively.  

Some features of such black hole production -- and its precursor physics at the Planck scale -- are highly dependent on the detailed model of the extra dimensional spacetime and brane-world realization of gauge forces and matter.  However, there are certain features which are fairly model independent and ``generic."  Typically these  dominate in an expansion in $M_D/E$, where $E$ is the collision energy.  This discussion will focus on such generic features.
  
 To describe the collider phenomenology of such black holes, one must understand their production and decay.  Many details of both were investigated in the papers \cite{GiTh,DiLa}, where it was in particular found that black hole production could be both {\it prolific} and have {\it outstanding signatures}.  After formation, a black hole decays in several phases\cite{GiTh}, termed {\it balding}, {\it spindown}, {\it Schwarzschild}, and {\it Planck}.  This paper will summarize these basic stages, and indicate improvements in understanding since these original papers.  Some parts of the story that are still missing are also indicated.
    
\section{Formation}

In a collision at energies $E\gg M_D$, one believes that many features of quantum gravity are adequately summarized by classical physics, with subleading effects suppressed by powers of $M_D/E$.  (For more discussion see \cite{Hsu,GiRy,GiL,GGM} and references therein.)  Classical production of black holes at high-energies was first studied by Penrose, who in unpublished work\cite{Penrose} showed that a collision at zero impact parameter forms a closed-trapped surface.  The ``area theorem" of general relativity then implies that this corresponds to formation of a true black hole.  
A more general construction for non-zero impact parameter and general dimension was later found in \cite{EaGi}, which confirmed the expectation \cite{Thorne} that black holes would form out to impact parameters comparable to the Schwarzschild radius,
\begin{equation}
R_S(E)= \left[{2(2\pi)^{D-4}E\over (D-2) \Omega_{D-2} M_D^{D-2}}\right]^{1/(D-3)}  \sim\quad  M_D^{-1} (E/M_D)^{1/(D-3)}\ .
\end{equation}
(Here $\Omega_{D-2}=2 \pi^{(D-1)/2}/\Gamma[(D-1)/2)]$ is the volume of the unit $(D-2)$-sphere.)
An important observation is that the trapped surface forms before the particles actually collide.  This is because the incident particles travel at essentially the speed of light, so that they can enter into a region smaller than $R_S$ before any signal of their arrival deforms the geometry in this region.  Thus, the trapped surface construction of \cite{Penrose,EaGi} can be made entirely in the well-understood pre-collision geometry, which consists of two Aichelburg-Sexl solutions.  

This construction then yields important features of production.  
Existence of a maximum impact parameter $b\sim R_S$ implies a cross section 
\begin{equation}\label{cross}
\sigma\sim \pi R_S^2\ .  
\end{equation}
Moreover, the area theorem states that classically the horizon area of the ultimate black hole must be greater than that of the original trapped surface, giving a lower bound on the mass of the black hole. The construction of \cite{EaGi} explicitly yields the trapped surface in $D=4$, but numerical calculation of its shape had to be performed in the  important work \cite{YoNa}, which then yielded more precise results for maximum impact parameters and lower bounds on black hole masses.   Another advance was the realization \cite{YoRy} that pushing the trapped surface construction as far into the future of the collision as possible before losing control of the dynamics yields modest improvements on the estimate of the cross section and a slight improvement on the lower bound on the mass.  

In the context of hadron colliders, the cross section (\ref{cross}) is the parton-level cross section, and parton distribution functions must be folded in to get production rates.  This question will be further discussed shortly, in the section on experimental expectations.  Some remaining uncertainties in production include the claim that charge effects can reduce cross sections\cite{YoMa}, as well as the fact that at LHC one doesn't really expect to be fully in the asymptotic regime $E\gg M_D$, so many other effects dependent on the detailed brane/standard model construction could make important corrections.  Improved understanding of these questions is needed.

\section{Decay}

\paragraph{\bf Balding phase}

When first formed, the black hole is very asymmetrical -- both its gravitational and other fields have very high multipole moments.  However, a stationary black hole has no ``hair" -- that is, to become stationary, the black hole must shed its higher multipole moments through a process termed {\it balding}.  In the $E\gg M_D$ limit, this is an essentially classical process, which takes place on a characteristic time scale $t\sim R_S$.  At the same time, the black hole should shed any charges carried by particles with large charge to mass ratios, through an analog of the Schwinger process; this is expected to include color and electric charge.

Thus, during balding, the black hole will emit gravitational and gauge radiation, and its charge.  The result will be a spinning, uncharged Kerr black hole.  The essentially classical balding process is very poorly understood at any level of detail -- this is a place for future improvement, perhaps via numerical methods.  An important question for the rest of the decay is the mass and spin of the resulting black hole, accounting for this ``inelasticity."  
Since balding (minus the charge loss) is a classical process, a lower bound on  mass is given by the trapped surface area as described above.   
Improved understanding of balding would also be useful to determine its contribution to experimental signatures.

\paragraph{\bf Spindown phase}

After balding the black hole is classically stable, but as predicted by Hawking, will continue to decay quantum-mechanically.  A rotating black hole first decays by preferentially shedding its angular momentum, in {\it spindown}.  The four-dimensional version of the process was studied long ago\cite{Page}.  To calculate details of this process, one must calculate higher-dimensional Hawking emission rates, which depend not just on the well known thermality of the Hawking radiation, but also on detailed ``gray-body" factors, that parametrize the effects of potential barriers for various modes near the black hole.  This is a hard problem.  

A basic parameter for higher dimensional black holes is the evaporation time scale,
\begin{equation}
t_H\sim M_D^{-1} (E/M_D)^{(D-1)/(D-3)}\ .
\end{equation}
Initial estimates for other parameters of spindown were made in \cite{GiTh}, based on extrapolations from the D=4 results\cite{Page}.  (This phase was not considered in \cite{DiLa}.)  There has been much subsequent work to actually calculate gray-body factors and provide more precise estimates (\cite{Kanti,IOP} and references therein), and we are nearing a more complete understanding of this phase.  The rate of spin loss vs. mass loss is qualitatively similar to that in D=4, though with the more refined calculations, \cite{IOP} claims that a black hole will lose over 50\% of its mass during spindown.  This is somewhat dependent on the rather arbitrary definition of the end of the spindown phase, but stresses the relevance of this phase.  It is clearly important to better understand the details of spindown, and experimental features of black hole decay will depend on these details (uncertainties over bulk graviton emission\cite{CNS,CCG} also remain).  One potentially important signature was pointed out in \cite{GiTh}: one might look for dipolar patterns characteristic of the black hole shedding its spin.

Detailed study of spindown has also begun to reveal other important parameters, like improved estimates of the relative rates of emission of vector, spinor, and scalar particles.  Apparently vector and spinor emission dominate.  The resulting ratios can yield detailed predictions for hadron to lepton ratios, etc., improving rough estimates of \cite{GiTh,DiLa}.  It is important for this program to be brought to completion, working out these and other specific signatures.

\paragraph{\bf Schwarzschild phase}

There has been greater focus on the details of this theoretically-simplest stage, where the black hole is non-rotating and decays via Hawking radiation.  Power spectra, relative emission rates of different particle species, etc. are largely determined by the fact that such a black hole emits essentially thermally, at an instantaneous temperature determined by its mass:
\begin{equation}
T_H= {D-3\over 4\pi R_S} \propto M^{-1/(D-3)}\ ,
\end{equation}
and this yields estimates of relative multiplicities of decay products\cite{GiTh,DiLa}.  
However, as with spindown, there can be important corrections due to gray-body factors, which can now be calculated\cite{Kanti,IOP} as opposed to estimated.  Modifications to the thermal multiplicities include suppression of low-energy gauge boson emission and other corrections.  

Future improvements, which seem close to being within reach, are a full study of the evolution of the black hole through {\it both} the spindown and Schwarzschild phases, properly incorporating gray-body factors, and integrating over the evolution, in order to give detailed predictions for critical features such as the energy spectrum, total and relative multiplicities, event shape parameters (such as angular distributions), etc.  

\paragraph{\bf Planck phase}

When the black hole reaches the mass scale $M\sim M_D$, all known physics breaks down and predictions cannot be made:  this is the fully quantum-gravitational realm.  Conversely, to  the extent to which we can see experimental signatures from this phase, we will learn about the dynamics of quantum gravity.  This is thus the most interesting phase.  One might in general expect emission of a few particles (or, if string theory is right, excited string states) with energies $E\sim M_D$ and high transverse momenta.  But who knows what surprises this phase might yield, if indeed it can be seen.

\section{Experimental expectations}

\paragraph{\bf Production rates}
The above discussion can now be assembled into a description of experimental expectations.  The first question is how low the Planck scale could be.  Different normalization conventions exist for this scale; for further discussion see \cite{Snowmass}.  The current treatment has adopted the convention of 
 \cite{GRW} (also used by the Particle Data Book), which differs from the definition of $M_p$ used in \cite{GiTh} by a factor:
 \begin{equation}
 M_p= 2^{1/(D-2)}M_D\ ;
 \end{equation}
this formula can be used to convert formulas in that paper.  Current bounds, quoted for example at SUSY 2007\cite{Duperrin} are dimension-dependent but lie around 
\begin{equation}\label{planckbd}
M_D\roughly> 1 {\rm TeV}\ ,
\end{equation}
with similar expectations for warped compactifications such as the toy model of \cite{RSone} or in string theory\cite{GKP,DoKa}.  Of course there are numerous theoretical difficulties in finding a complete model with such a scale, but we will take the viewpoint that the true Planck scale could lie in this vicinity and explore the consequences.

For a given Planck mass, the next question is for what collision energies the $M_D/E$ expansion begins to yield reliable results.  There are a number of possible criteria, many discussed in \cite{GiTh}; this reference advocated that one particularly useful criterion is that the entropy of the black hole be large, so that a thermal approximation begins to make sense.  A non-rotating hole of mass $M$ has entropy
\begin{equation}
S_{BH}= \left[2M/ (D-2) M_D \right]^{(D-2)/(D-3)}(2\pi)^{(2D-7)/(D-3)} \Omega_{D-2}^{-1/(D-3)}\ .
\end{equation}
 For example, for the representative values $D=6$ and $D=10$, a black hole with mass 
\begin{equation}\label{critmass}
M=5 M_D
\end{equation}
has entropy  $S_{BH}\simeq 24$. 
Thus, a plausible threshold for semiclassical black hole effects could lie around $5M_D$.

Next consider production rates.  The rough cross section (\ref{cross}) can be improved by the results of \cite{EaGi,YoNa,YoRy}.  One must account not only for the modification of the cross-section, but also for the fact that the inelasticity grows at increasing impact parameter, so one can drop below a threshold such as 
(\ref{critmass}) at sufficiently large impact parameter.  The trapped-surface calculations of \cite{EaGi,YoNa,YoRy} provide {\it lower} bounds on the mass of the black hole.   In four dimensions and with zero impact parameter, such a trapped-surface bound gives $M\geq E/\sqrt2\simeq.71E$.  However, subsequent estimates of \cite{Death} raise the estimated mass to $M\approx .84 E$.  While inelasticity was considered in \cite{GiTh}, its effects were underestimated based on these then state-of-the-art four-dimensional estimates -- \cite{EaGi} revealed that  inelasticity increases with dimension.
Indeed, in the higher-dimensional situation, inspection of the figures in \cite{YoRy} show that the trapped-surface bounds are lower in higher dimensions.  Analogs of the estimates \cite{Death} don't yet exist.\footnote{Recent work\cite{MeRa} pessimistically takes such lower bounds to represent the actual mass.}

The trapped surface lower bounds of \cite{EaGi,YoNa,YoRy} can be roughly parametrized by a curve of the form
\begin{equation}
M= .5 E\ , b<.5 R_S\quad ;\quad M= 0\ , b>.5 R_S\ .
\end{equation}
This parametrization follows from fig.~10 of \cite{YoRy}, for $D=10$, with the cutoff taking into account a threshold of the form (\ref{critmass}).  For lower $D$ this parametrization represents a significant {\it underestimate}.  Moreover, as exemplified by the four-dimensional estimates \cite{Death}, one expects that the actual mass should be higher.  Thus, for the purposes of making estimates, we will consider the following two ``production efficiency curves:"
\begin{eqnarray}\label{prodeff}
 (I):\quad &&M= .6 E\ , b<.5 R_S\quad ;\quad M= 0\ , b>.5 R_S\\
(II):\quad &&M= .7 E\ , b<.5 R_S\quad ;\quad M= 0\ , b>.5 R_S\ \ .
\end{eqnarray}
These very rough estimates have corrections of both signs; more precise calculations should be part of future work.  

With a black hole threshold of 5 TeV, the curves (\ref{prodeff}) correspond to a minimum parton CM energy of 8.3 and 7.1 TeV, respectively.  By these energies, the parton distribution functions have fallen significantly.  Total cross sections are readily estimated; the parameterizations (I) and (II) give us a parton cross section $\sigma = \pi R_S^2/4$ for partons above the critical energy.  From these cross sections, calculations using the CTEQ6M parton distribution functions\footnote{I gratefully acknowledge Tom Rizzo's calculation of these cross sections.} yield:
\begin{equation}
(I): \sigma = 1.8 \times 10^2 fb\quad ;\quad (II): \sigma = 1.8 \times 10^3 fb\ .
\end{equation}
These (still rough) estimates are significantly lower than the original estimates of \cite{GiTh,DiLa}, due to the substantial inelasticity revealed in the subsequent work \cite{EaGi,YoNa,YoRy}.  This illustrates the great sensitivity of cross sections to the production efficiency curves, motivating their better understanding.  Still, the estimates (I), (II) yield respectable rates:  at the nominal LHC luminosity of $10^{34}/cm^2 s$, one black hole every ten minutes for (I), and one every minute for (II).  However, such considerations also illustrate a narrowed parameter window where black hole production is possible, as has been stressed in \cite{AFGS,MeRa}.  

\paragraph{\bf Signatures}
To determine explicit and quantitative signatures, one needs to finalize the full integrated description of the evolution through the spindown and the Schwarzschild phases, including the important effects of gray-body factors, as described above.  These results could for example be used in conjunction with event generators.  Three such generators exist: TRUENOIR, CHARYBDIS, and Catfish.  However, none of them properly account for spindown, which is an important feature of the decay\cite{GiTh,IOP}.  Thus significant work remains to be done.

Nonetheless, some rather striking qualitative signatures can be inferred\cite{GiTh,DiLa}.  These include

\begin{itemize}
\item potentially large cross sections, approaching $10^3$ fb or more (also an increase of cross section with energy, according to (\ref{cross}), which will be hard to see at LHC)
\item relatively high sphericity;
\item high multiplicity ($\propto S_{BH}$) of primary particles produced;
\item hard transverse leptons and jets, in significant numbers;
\item approximately thermally determined ratios of species;
\item angular distributions characterizing the spindown phase;
\item suppression of highest-energy jets\cite{BaFi,GiTh,HBGHSS};
\item decrease of primary final state lepton/parton energy with total event transverse energy, resulting from decreasing Hawking temperature with mass.
\end{itemize}

Some of these signatures may be most visible through multi-event statistics, particularly near threshold, where individual event multiplicities may not yield good statistics.  (Indeed, \cite{AFGS} estimates multiplicity $\langle N\rangle \approx S_{BH}/3$ -- which should now be checked with improved gray-body calculations.)

\paragraph{\bf Cosmic ray production}

Immediately after \cite{GiTh,DiLa}, it was realized that black holes should be produced in collisions of hadronic ultrahigh-energy cosmic rays with the atmosphere, whose CM energies range upwards of 100 TeV, but that the rate would be overwhelmed by QCD processes and thus be too low to be observable\cite{Snowmass}.  However, one expects such cosmic rays to interact with the microwave background and produce a flux of ultrahigh-energy GZK cosmic ray neutrinos.  These could produce an observable rate of black hole events\cite{DGRT,FeSh,AnGo,Snowmass}.  In particular, the possibility of seeing such events at cosmic ray facilities such as AMANDA, Auger, IceCube, or others has now been widely investigated.

In particular, \cite{AFGS} argued that non-observation of such events with five years of Auger data could yield a bound $M_D\roughly>2$ TeV, and rule out production of semiclassical black holes at LHC.  It will be interesting to keep an eye on this data.  There are possible uncertainties in such a discussion; examples include arguments that neutrino fluxes can be suppressed in large extra dimension scenarios\cite{LMR} and a proposal\cite{DSS} that $\nu p\rightarrow BH$ events could be suppressed by the physics preserving baryon number, and uncertainties over primary composition\cite{AnGoetal}.  So, while we may be lucky enough to see hints from Auger, ultimately we should wait to see what LHC brings.  

\section{Theoretical issues}

Whether or not black hole production lies within reach depends on whether nature is able to realize a TeV-scale gravity scenario.  Construction of such scenarios has been very challenging, but there has been a lot of progress and ongoing developments too numerous to review here.  Ultimately, experiment will be the jury.

However, even if black hole production is not experimentally accessible, it is an extremely important theoretical problem, as it forces confrontation with our most profound theoretical issues.  Notable among these is the black hole information paradox.\footnote{For reviews see \cite{SGTr,Astro}.} The basic statement of this paradox is that consideration of the fate of quantum information in the context of evaporating black holes apparently forces us to abandon a cherished principle of physics.  The possibilities include abandoning unitary quantum-mechanical evolution, as originally suggested by Hawking\cite{Hawkunc}, with the apparent consequent disastrous abandonment of energy conservation\cite{BPS}; abandoning stability, as implied by a black hole remnant scenario, or abandoning macroscopic locality, in order that information can escape a black hole in Hawking radiation.  

\paragraph{\bf Nonlocality}
There is not space here for a full discussion of the situation, but it is worth summarizing some highlights.  First, an increasingly widespread belief, particularly in the string theory community, is that locality is not correct, and information escapes during evaporation of a black hole.  There has been a great deal of discussion of this in this in the context of ideas such as the ``holographic principle."  However, if nonlocality is the answer to the paradox, it is very important to answer some basic questions, such as:  1) what is the mechanism for locality violation; 2) how is Hawking's original argument for information loss avoided and 3) in what contexts do semiclassical general relativity(GR) together with local quantum field theory(QFT) fail?  One thing that would be very interesting is to properly establish the boundaries of a {\it correspondence limit} for the new physics, whether string theory or something else, that replaces GR+QFT.

An obvious possibility,  widely considered by string theorists, is that such nonlocality has as its origin the nonlocal extendedness of strings.  This possibility can be investigated precisely in the context of high-energy collisions.  The result is recent arguments\cite{GiL,GGM} that high-energy scattering evidences nonlocal behavior, but {\it there is no evidence of a role for string extendedness}.  Instead, it has been proposed\cite{Nonlocp} that nonlocality could emerge as an effect intrinsic to the non-perturbative dynamics of quantum gravity.  

An important question is how precisely one can test the notion of locality.  In QFT, locality is encoded in commutativity of local observables at  spacelike separations.  Diffeomorphism invariance, the gauge symmetry of gravity, implies that there are no such local observables.  One can instead formulate certain ``proto-local" observables\cite{GMH,GaGi}, which are diffeomorphism-invariant objects that approximately reduce to local observables in appropriate limits.  However, limits on when this reduction is possible include those from situations where gravity becomes strong, such as in formation of a black hole.  Another test of locality is in the asymptotic behavior of scattering amplitudes at high energies.  Behavior such as (\ref{cross}), and Hawking (thermal) decay amplitudes, violate the known useful criteria for locality such as the Froissart\cite{froiss} and possibly Cerulus-Martin\cite{CeMa} bounds.  Together, these statements suggest that there is no precise notion of locality in quantum gravity, and support the above statement that its breakdown is associated with non-perturbative gravitational dynamics.

This guides us towards parametrizing the correspondence limit where GR+QFT are expected to break down; one parametrization is in the form of  the ``locality bounds" of \cite{Nonlocp,Locbd,GiMa}.  For example, one expects a Fock space description of a two-particle state consisting of minimum-uncertainty wavepackets with approximate positions $x,y$ and momenta $p,q$ to only be a good approximation to a complete quantum description when
\begin{equation}
|x-y|^{D-3} > G|p+q|\ ,
\end{equation}
where $G$ is a constant proportional to the $D$-dimensional Newton constant.

Another important question is how these ideas relate to the notion that there should be a good effective field theory description of the interior of a large black hole, which lies at the heart of Hawking's argument for information loss.  It may be that new nonlocal effects in black holes are as invisible to the framework of GR+QFT as quantum effects are to a classical description of the atom.  However, there are some initial indications otherwise\cite{QBHB}.  In particular, a careful attempt to justify Hawking's derivation of information loss in an expansion in $M_D/E$ reveals an apparent breakdown of the calculation, at the long time scales where we would expect information to begin to escape a black hole; this is a proposed resolution to the information paradox.  Whether or not this breakdown of a perturbative derivation of information loss is the ultimate rationale, a proposal is that {\it information escapes a black hole through nonperturbative gravitational effects that only become important at long times}\cite{Nonlocp,QBHB,Arkani}.  These would not respect usual notions of locality.

These ideas should also be relevant in the cosmological context\cite{Arkani,QBHB,GiMa}, and may alter the picture that has led to conundrums of modern cosmology, such as the landscape and Boltzman brains.  Such issues are under active investigation.  

Without prejudicing the discussion in favor of string theory, an ultimate question in this direction is what are the basic principles of such a ``nonlocal mechanics," providing the complete nonperturbative description of quantum gravity.

%
%
%
%

%
%
%
%
%
%


\begin{theacknowledgments}
  I thank the organizers of PASCOS for the invitation to a very stimulating conference. I am particularly grateful to T. Rizzo for conversations and  cross-section computations, and to A. Shapere for conversations. This work  was supported in part by the DOE under Contract DE-FG02-91ER40618, and by grant RFPI-06-18 from the
Foundational Questions Institute (fqxi.org).
\end{theacknowledgments}

\bibliographystyle{aipproc}   

\begin{thebibliography}{9}

\bibitem{Kantir}
  P.~Kanti,
  ``Black holes in theories with large extra dimensions: A review,''
  Int.\ J.\ Mod.\ Phys.\  A {\bf 19}, 4899 (2004)
  [arXiv:hep-ph/0402168].
  
 \bibitem{Webber}
  B.~Webber,
  ``Black holes at accelerators,''
{\it In the Proceedings of 33rd 
SLAC Summer Institute on Particle Physics (SSI 2005): Gravity in the Quantum World and the Cosmos, Menlo Park,
California, 25 Jul - 5 Aug 2005, pp T030}
  [arXiv:hep-ph/0511128].

\bibitem{KBS}
  B.~Koch, M.~Bleicher and H.~Stoecker,
  ``Black holes at LHC?,''
  J.\ Phys.\ G {\bf 34}, S535 (2007)
  [arXiv:hep-ph/0702187].


\bibitem{ADD}
  N.~Arkani-Hamed, S.~Dimopoulos and G.~R.~Dvali,
  ``The hierarchy problem and new dimensions at a millimeter,''
  Phys.\ Lett.\  B {\bf 429}, 263 (1998)
  [arXiv:hep-ph/9803315].

\bibitem{AADD}
  I.~Antoniadis, N.~Arkani-Hamed, S.~Dimopoulos and G.~R.~Dvali,
  ``New dimensions at a millimeter to a Fermi and superstrings at a TeV,''
  Phys.\ Lett.\  B {\bf 436}, 257 (1998)
  [arXiv:hep-ph/9804398].

\bibitem{RSone}
  L.~Randall and R.~Sundrum,
  ``A large mass hierarchy from a small extra dimension,''
  Phys.\ Rev.\ Lett.\  {\bf 83}, 3370 (1999)
  [arXiv:hep-ph/9905221].

\bibitem{GKP}
  S.~B.~Giddings, S.~Kachru and J.~Polchinski,
  ``Hierarchies from fluxes in string compactifications,''
  Phys.\ Rev.\  D {\bf 66}, 106006 (2002)
  [arXiv:hep-th/0105097].

\bibitem{DoKa}
  M.~R.~Douglas and S.~Kachru,
  ``Flux compactification,''
  Rev.\ Mod.\ Phys.\  {\bf 79}, 733 (2007)
  [arXiv:hep-th/0610102].

\bibitem{BaFi}
  T.~Banks and W.~Fischler,
  ``A model for high energy scattering in quantum gravity,''
  arXiv:hep-th/9906038.

\bibitem{GiKa}
  S.~B.~Giddings and E.~Katz,
  ``Effective theories and black hole production in warped
  compactifications,''
  J.\ Math.\ Phys.\  {\bf 42}, 3082 (2001)
  [arXiv:hep-th/0009176].

\bibitem{GiTh}
  S.~B.~Giddings and S.~D.~Thomas,
  ``High energy colliders as black hole factories: The end of short  distance
  physics,''
  Phys.\ Rev.\  D {\bf 65}, 056010 (2002)
  [arXiv:hep-ph/0106219].

\bibitem{DiLa}
  S.~Dimopoulos and G.~L.~Landsberg,
  ``Black holes at the LHC,''
  Phys.\ Rev.\ Lett.\  {\bf 87}, 161602 (2001)
  [arXiv:hep-ph/0106295].

\bibitem{Hsu}
  S.~D.~H.~Hsu,
  ``Quantum production of black holes,''
  Phys.\ Lett.\  B {\bf 555}, 92 (2003)
  [arXiv:hep-ph/0203154].

\bibitem{GiRy}
  S.~B.~Giddings and V.~S.~Rychkov,
  ``Black holes from colliding wavepackets,''
  Phys.\ Rev.\  D {\bf 70}, 104026 (2004)
  [arXiv:hep-th/0409131].

\bibitem{GiL}
  S.~B.~Giddings,
  ``Locality in quantum gravity and string theory,''
  Phys.\ Rev.\  D {\bf 74}, 106006 (2006)
  [arXiv:hep-th/0604072].

\bibitem{GGM}
  S.~B.~Giddings, D.~J.~Gross and A.~Maharana,
  ``Gravitational effects in ultrahigh-energy string scattering,''
  arXiv:0705.1816 [hep-th].

\bibitem{Penrose} R. Penrose, unpublished (1974).

\bibitem{EaGi}
  D.~M.~Eardley and S.~B.~Giddings,
  ``Classical black hole production in high-energy collisions,''
  Phys.\ Rev.\  D {\bf 66}, 044011 (2002)
  [arXiv:gr-qc/0201034].

\bibitem{Thorne}
  K.~S.~Thorne,
  ``Nonspherical Gravitational Collapse: A Short Review,''
{\it  In *J R Klauder, Magic Without Magic*, San Francisco 1972, 231-258}

\bibitem{YoNa}
  H.~Yoshino and Y.~Nambu,
  ``Black hole formation in the grazing collision of high-energy particles,''
  Phys.\ Rev.\  D {\bf 67}, 024009 (2003)
  [arXiv:gr-qc/0209003].

\bibitem{YoRy}
  H.~Yoshino and V.~S.~Rychkov,
  ``Improved analysis of black hole formation in high-energy particle
  collisions,''
  Phys.\ Rev.\  D {\bf 71}, 104028 (2005)
  [arXiv:hep-th/0503171].

\bibitem{YoMa}
  H.~Yoshino and R.~B.~Mann,
  ``Black hole formation in the head-on collision of ultrarelativistic
  charges,''
  Phys.\ Rev.\  D {\bf 74}, 044003 (2006)
  [arXiv:gr-qc/0605131].

\bibitem{Page}
  D.~N.~Page,
  ``Particle Emission Rates From A Black Hole. 2. Massless Particles From A
  Rotating Hole,''
  Phys.\ Rev.\  D {\bf 14}, 3260 (1976).
  
\bibitem{Kanti}
  G.~Duffy, C.~Harris, P.~Kanti and E.~Winstanley,
  ``Brane decay of a (4+n)-dimensional rotating black hole: Spin-0
  particles,''
  JHEP {\bf 0509}, 049 (2005)
  [arXiv:hep-th/0507274]; 
 M.~Casals, P.~Kanti and E.~Winstanley,
  ``Brane decay of a (4+n)-dimensional rotating black hole. II: Spin-1
  particles,''
  JHEP {\bf 0602}, 051 (2006)
  [arXiv:hep-th/0511163];
M.~Casals, S.~R.~Dolan, P.~Kanti and E.~Winstanley,
  ``Brane decay of a (4+n)-dimensional rotating black hole. III: Spin-1/2
  particles,''
  JHEP {\bf 0703}, 019 (2007)
  [arXiv:hep-th/0608193].


\bibitem{IOP}
D.~Ida, K.~y.~Oda and S.~C.~Park,
  ``Rotating black holes at future colliders: Greybody factors for brane
  fields,''
  Phys.\ Rev.\  D {\bf 67}, 064025 (2003)
  [Erratum-ibid.\  D {\bf 69}, 049901 (2004)]
  [arXiv:hep-th/0212108];
  ``Rotating black holes at future colliders. II: Anisotropic scalar field
  emission,''
  Phys.\ Rev.\  D {\bf 71}, 124039 (2005)
  [arXiv:hep-th/0503052];
   ``Rotating black holes at future colliders. III: Determination of black  hole
  evolution,''
  Phys.\ Rev.\  D {\bf 73}, 124022 (2006)
  [arXiv:hep-th/0602188].

\bibitem{CNS}
  A.~S.~Cornell, W.~Naylor and M.~Sasaki,
  ``Graviton emission from a higher-dimensional black hole,''
  JHEP {\bf 0602}, 012 (2006)
  [arXiv:hep-th/0510009].

\bibitem{CCG}
  V.~Cardoso, M.~Cavaglia and L.~Gualtieri,
  ``Hawking emission of gravitons in higher dimensions: Non-rotating black
  holes,''
  JHEP {\bf 0602}, 021 (2006)
  [arXiv:hep-th/0512116].


\bibitem{Snowmass}
  S.~B.~Giddings,
  ``Black hole production in TeV-scale gravity, and the future of high  energy
  physics,''
{\it In the Proceedings of APS / DPF / DPB Summer Study on the Future of Particle Physics (Snowmass 2001), Snowmass, Colorado, 30 Jun - 21 Jul
2001, pp P328}, ed. N.~Graf.
  [arXiv:hep-ph/0110127].

\bibitem{GRW}
  G.~F.~Giudice, R.~Rattazzi and J.~D.~Wells,
  ``Quantum gravity and extra dimensions at high-energy colliders,''
  Nucl.\ Phys.\  B {\bf 544}, 3 (1999)
  [arXiv:hep-ph/9811291].

\bibitem{Duperrin}
A. Duperrin, ``Direct searches for Higgs and Beyond the SM at the Tevatron,"  talk at SUSY 2007, Karlsruhe, Germany.

\bibitem{Death}
  P.~D.~D'Eath,
  ``Gravitational radiation in high speed black hole collisions,''
  Class.\ Quant.\ Grav.\  {\bf 10}, S207 (1993).

\bibitem{MeRa}
  P.~Meade and L.~Randall,
  ``Black Holes and Quantum Gravity at the LHC,''
  arXiv:0708.3017 [hep-ph].

\bibitem{AFGS}
  L.~A.~Anchordoqui, J.~L.~Feng, H.~Goldberg and A.~D.~Shapere,
 ``Inelastic black hole production and large extra dimensions,''
  Phys.\ Lett.\  B {\bf 594}, 363 (2004)
  [arXiv:hep-ph/0311365].


\bibitem{HBGHSS}
  S.~Hofmann, M.~Bleicher, L.~Gerland, S.~Hossenfelder, S.~Schwabe and H.~Stoecker,
  ``Suppression of high-P(T) jets as a signal for large extra dimensions  and
  new estimates of lifetimes for meta stable micro black holes:  From the early
  universe to future colliders,''
  arXiv:hep-ph/0111052.

\bibitem{DGRT}
D. Dorfan, S.B. Giddings, T. Rizzo, and S. Thomas, unpublished (2001).

\bibitem{FeSh}
  J.~L.~Feng and A.~D.~Shapere,
  ``Black hole production by cosmic rays,''
  Phys.\ Rev.\ Lett.\  {\bf 88}, 021303 (2002)
  [arXiv:hep-ph/0109106].

\bibitem{AnGo}
  L.~Anchordoqui and H.~Goldberg,
  ``Experimental signature for black hole production in neutrino air
  showers,''
  Phys.\ Rev.\  D {\bf 65}, 047502 (2002)
  [arXiv:hep-ph/0109242].





\bibitem{LMR}
  J.~Lykken, O.~Mena and S.~Razzaque,
 ``Ultrahigh-energy neutrino flux as a probe of large extra-dimensions,''
  arXiv:0705.2029 [hep-ph].

\bibitem{DSS}
  D.~Stojkovic and G.~D.~Starkman,
  ``Why black hole production in scattering of cosmic ray neutrinos is
  generically suppressed,''
  Phys.\ Rev.\ Lett.\  {\bf 96}, 041303 (2006)
  [arXiv:hep-ph/0505112].
  
  
\bibitem{AnGoetal}
  L.~A.~Anchordoqui, H.~Goldberg, D.~Hooper, S.~Sarkar and A.~M.~Taylor,
  ``Predictions for the Cosmogenic Neutrino Flux in Light of New Data from the
 Pierre Auger Observatory,''
  arXiv:0709.0734 [astro-ph].

\bibitem{SGTr}
  S.~B.~Giddings,
  ``Quantum mechanics of black holes,''
  arXiv:hep-th/9412138;
``The Black hole information paradox,''
  arXiv:hep-th/9508151.

\bibitem{Astro}
  A.~Strominger,
  ``Les Houches lectures on black holes,''
  arXiv:hep-th/9501071.

\bibitem{Hawkunc}
  S.~W.~Hawking,
  ``Breakdown Of Predictability In Gravitational Collapse,''
  Phys.\ Rev.\  D {\bf 14}, 2460 (1976).

\bibitem{BPS}
  T.~Banks, L.~Susskind and M.~E.~Peskin,
  ``Difficulties For The Evolution Of Pure States Into Mixed States,''
  Nucl.\ Phys.\  B {\bf 244}, 125 (1984).

\bibitem{Nonlocp}
  S.~B.~Giddings,
  ``Black hole information, unitarity, and nonlocality,''
  Phys.\ Rev.\  D {\bf 74}, 106005 (2006)
  [arXiv:hep-th/0605196];
``(Non)perturbative gravity, nonlocality, and nice slices,''
  Phys.\ Rev.\  D {\bf 74}, 106009 (2006)
  [arXiv:hep-th/0606146].

\bibitem{GMH}
  S.~B.~Giddings, D.~Marolf and J.~B.~Hartle,
  ``Observables in effective gravity,''
  Phys.\ Rev.\  D {\bf 74}, 064018 (2006)
  [arXiv:hep-th/0512200].

\bibitem{GaGi}
  M.~Gary and S.~B.~Giddings,
 ``Relational observables in 2d quantum gravity,''
  Phys.\ Rev.\  D {\bf 75}, 104007 (2007)
  [arXiv:hep-th/0612191].


\bibitem{froiss}
  M.~Froissart,
  ``Asymptotic behavior and subtractions in the Mandelstam representation,''
  Phys.\ Rev.\  {\bf 123} (1961) 1053.

\bibitem{CeMa}
F. Cerulus and A. Martin, ``A lower bound for large angle elastic scattering at high
energies," Phys. Lett. {\bf 8} (1964) 80.

\bibitem{Locbd}
  S.~B.~Giddings and M.~Lippert,
  ``Precursors, black holes, and a locality bound,''
  Phys.\ Rev.\  D {\bf 65}, 024006 (2002)
  [arXiv:hep-th/0103231];
``The information paradox and the locality bound,''
  Phys.\ Rev.\  D {\bf 69}, 124019 (2004)
  [arXiv:hep-th/0402073].


\bibitem{GiMa}
  S.~B.~Giddings and D.~Marolf,
  ``A global picture of quantum de Sitter space,''
  arXiv:0705.1178 [hep-th].

\bibitem{QBHB}
  S.~B.~Giddings,
 ``Quantization in black hole backgrounds,''
  arXiv:hep-th/0703116, to appear in Phys. Rev. D.

\bibitem{Arkani}
  N.~Arkani-Hamed, S.~Dubovsky, A.~Nicolis, E.~Trincherini and G.~Villadoro,
  ``A Measure of de Sitter Entropy and Eternal Inflation,''
  JHEP {\bf 0705}, 055 (2007)
  [arXiv:0704.1814 [hep-th]].

\end{thebibliography}

\end{document}